\documentclass[aps,floatfix,superscriptaddress,notitlepage,nofootinbib]{revtex4-1}
\usepackage[utf8]{inputenc}
\usepackage{tikz}
\usepackage{amsmath,amssymb,amsfonts,graphics,graphicx,dcolumn,bm,enumerate}
\usepackage{comment,natbib,appendix}
\usepackage{multirow,color}
\usepackage{chngpage}
\usepackage{afterpage}
\usepackage{xcolor}
\usepackage{amsthm}
\usepackage{natbib}
\usepackage{hyperref}
\usepackage[margin=0.8in]{geometry}
\usepackage{epstopdf}
\usepackage{float}
\usepackage{soul}

\newcommand{\isi}
{\affiliation{Economic Research Unit, Indian Statistical Institute, Kolkata 700108, India.}}

\begin{document}

\title{Role of Neighbouring Wealth Preference in Kinetic Exchange model of market}

\author{Suchismita Banerjee}
\email[Email: ]{suchib.1993@gmail.com}
\thanks{corresponding author}
\isi

\begin{abstract} 
The kinetic exchange model has gained popularity in the field of statistical mechanics for investigating wealth interaction. 
Traditionally, kinetic exchange models have been studied without considering preferential interactions. 
However, in this study, we introduce two types of preferential interactions to explore wealth dynamics and its associated distributions.
In the first preference, one agent is randomly selected, while the other agent is chosen randomly with wealth just above or below the first agent. 
Through this preference, we observe the emergence of a quasi-oligarchic society, where the majority of the wealth cycles around the hand of very few agents.
For the second preference, we impose a constraint on the difference in pre-interaction wealth between the two agents. This preference leads to the segregation of society into two distinct economic classes.
To investigate these phenomena, we conducted extensive Monte Carlo simulations, enabling us to characterize the behavior of wealth distributions in these two scenarios. Our findings shed light on the dynamics of wealth accumulation and distribution within preferential interactions in the context of the kinetic exchange model.
\end{abstract}

\maketitle

\section*{Introduction} For 90 to 95 percent of people in any country seen to fall in an exponentially decaying both wealth and income distributions \cite{Tao_2019},\cite{Dragulescu_2001},\cite{Dragulescu_2003},\cite{Silva_2005}. That is, for some parameter $w_0$, the number density of people $P(w)$ with wealth w is proportional to $exp[-w/w_0]$. This pattern holds true for income statistics from all 67 nations investigated \cite{Tao_2019}. We can connect an inverse temperature with each economy $\beta = w_0$ because these wealth distributions are exponential. As the population of the United States becomes wealthier, the temperature has been observed to progressively rise over time \cite{Silva_2005}.

On the other hand, the income and wealth distributions of the wealthiest individuals of society are not exponential, but rather follow a power law: $P(w) \propto w^{-\alpha}$, with $\alpha \simeq 2$ across various data sets \cite{Dragulescu_2001},\cite{Souma_2001}, \cite{Souma_2005}, \cite{Silva_2005}. This empirical occurrence is known as Pareto's law, which was discovered by Pareto in 1897 while investigating the wealth distribution of wealthy Italian landowners \cite{Flux_1897}. The wealth of the richest persons in the United States is closely associated with the S\&P500, according to reference \cite{Dragulescu_2001}.

Agent-based modelling has a lot of promise in terms of improving our understanding of economics \cite{Farmer_2009},\cite{Arthur_2006},\cite{Gualdi_2013},\cite{Richiardi_2017}. Many assumptions concerning the functioning of free markets are based on classical economic theory. Economies have traditionally been thought to be in equilibrium, completely efficient, and constant over time. Agent-based modelling has the ability to provide insights not expected by conventional economics because it addresses the fundamental dynamics underpinning the broader economy. Agent-based modelling, rather than explicitly modelling an economy as a whole, is concerned with the behaviour of individual interacting agents. The collective behaviour of these numerous agents, in turn, models the economy.

A large number of agents, each with a given amount of wealth w, interact with other agents to update their wealth in an agent-based asset exchange model. The resulting wealth distribution across the system can be evaluated. Typically, the trade rule is stochastic, meaning that agents give and take wealth from one other at random. The system's final stationary wealth distribution, if one exists, is usually determined by the specific trade mechanism used between agents.

Agent-based models are frequently used to simulate real-world wealth distribution, but they may also be used to simulate numerous macroeconomic phenomena such as unemployment \cite{Gualdi_2013} and the impact of social programs \cite{Cardoso_2020}, and they are particularly helpful as a policy-guiding tool for monetary authorities. Because a researcher can make a variety of seemingly arbitrary choices when developing even a simple agent-based model, there are many variations of comparable models that provide qualitatively different outcomes.

Physicists have examined a variety of simple exchange rules, including random additive and multiplicative exchange. As $t\longrightarrow \infty$, some unfortunate agents will have negative total wealth due to simple additive trade, in which agents lose or earn a small, fixed quantity of money from each other at random. When these agents are removed from the economy after reaching zero wealth, the resultant wealth distribution is Gaussian \cite{Ispolatov_1998}. This distribution is not realistic, as previously stated.

Many economists argue that multiplicative exchange, in which each agent's wealth traded is proportional to their present wealth, is a superior model of economic activity. When the quantity of wealth traded between two agents is proportional to the wealth of the losing agent, the resulting wealth distribution is exponential for a wide range of parameters \cite{Ispolatov_1998}, which is a more realistic result \cite{Tao_2019},\cite{Silva_2005}. Mean-field theory was used to solve a comparable random, multiplicative agent-based model, in which victorious agents earn a fraction of the losing agent's wealth, but the agents' wealth is also influenced by additive Gaussian noise \cite{Bouchaud_2000}. For the wealthiest agents, the resulting wealth distribution is a power-law. This feature is in line with the real-world Pareto's law that was previously discussed. Even when a simple tax mechanism was added to the random, multiplicative model, the same distribution was shown to hold \cite{Bouchaud_2000}.

Pellicer-Lostao and Lopez-Ruiz \cite{Pellicer-Lostao_2011} proposed an interesting multiplicative agent-based model in which the winners and losers of an exchange are selected chaotically according to a two-dimensional chaotic system: the logistic bimap, rather than randomly (as in most agent-based models). The logistic bimap is a deterministic method of picking winning and losing pairs of agents to swap money. This chaotic system allows the researchers to modify the level of symmetry for each pair of agents; for low symmetry, a single agent is more likely to always win or always lose the exchanges, but for high symmetry, each exchange is effectively random. Pellicer-Lostao and Lopez-Ruiz discovered that when the amount of symmetry of the logistic bimap is reduced, the resultant stationary agent wealth distribution shifts from an exponential to a power law. In other words, the wealth distribution is exponential when the winner of a transaction is selected effectively randomly (high symmetry), but it is a power law when agents have varied chances of winning or losing their deals. This surprising conclusion shows that the poorest members of the economy (the bottom 97\%) have their wealth determined at random, whereas the wealthiest members of the economy ( the top 3\%) have their wealth decided by a constant ability to win (or lose) transactions.Ref.\cite{Sinha_2005} discusses a similar observation regarding the difference between rich and poor agents.

There have also been attempts to predict economic growth without taking into account the impact of agent exchange \cite{Vallejos_2017},\cite{Evers_2017}. Vallejos et al. used a discrete, stochastic growth and distribution algorithm in Ref.\cite{Vallejos_2017}, in which agent i receives new money with probability, 
\begin{equation}
    \psi_i = \frac{w_{i}^{\lambda}}{\sum_{j} w_{j}^{\lambda}},
\end{equation}
with the parameter $\lambda$ determining how much more likely the rich are to obtain new wealth than the poor. This rule is similar to the one proposed by Kang et al. in Ref.\cite{Liu_2013}. Vallejos et al. show that the exponential and power-law wealth distributions of real economies may be reproduced using only Eq.(1).

Evers et al.\cite{Evers_2017} use a one-dimensional lattice of economic agents to describe the globalisation process. These agents do not swap wealth with one another, but instead earn a ``wage" proportional to the average agent's wealth at each time step. Evers et al. rescale all the agents' wealth in the neighbourhood of the agent who receives a wage to keep the system's wealth from rising exponentially. In other words, getting new income in the form of a wage rescales the wealth of the agent and its neighbours so that the total quantity of wealth in each neighbourhood remains constant over time. The model's proxy for globalisation is the size of this neighbourhood, measured in units of lattice spacing. The results were found to be compatible with Kuznets' hypothesis regarding globalisation and inequality \cite{Kuznets_1963} as the size of this rescaling neighbourhood was increased. The process of globalisation first increases inequality between agents; but, as globalisation expands to include all agents in the system, inequality between agents decreases. Kuznets's "U-shaped" hypothesis is a popular name for this idea. 

With one of our preferential models, we observe the emergence of a Quasi-oligarchic society. 
Such kind of observation has been reported earlier as well in the context of Kinetic Exchange Model \cite{Chakrabarti}, \cite{Boghusian}.
However in the earlier case, the interaction is observed to get seized after few cycles and the wealth is observed to be concentrated in only one agent hands, while in our case we do not see such concentration of wealth in one agent's hand but it is observed to revolving around the hands of very few number of agents.
Moreover, for our case the interaction never stops as it was for the earlier cases.


\section*{Models for preferential interactions}

In this section, we describe the kinetic wealth exchange models (KWEM) which are adopted for this work.
These KWEMs consider a community of N agents with equal possession of wealth and randomly select two agents at any instant of time.
These two agents subsequently interact following a set of wealth conservative rules with each other. 
In this work, we consider two different selection criteria for choosing two agents, randomly.
We term the first model as Nearest Neighbour Wealth Model (NNWM) and the other one as Delta Distance Wealth Model (DDWM).
Below we describe the KWEMs that we consider in this work for completeness, and then we introduce the two preferences, studying whose effect is the main objective of this work. 

To begin with, we consider the celebrated kinetic exchange interaction proposed by \cite{Dragulescu_2000}, where the interaction rules between two random chosen agents are the following:
\begin{eqnarray}\label{eq:dragulescu}
\nonumber
    v^{*} &=& \eta(v+w) \\
    w^{*} &=& (1-\eta)(v+w) 
\end{eqnarray}
where, $\{v,w\}$ and $\{v^{*},w^{*}\}$ are the wealth of the randomly selected agents before and after contact at an instantaneous time step, respectively. 
$\eta\in\{0,1\}$ is a random number chosen from an uniform distribution. 
This interaction is known to give rise to an steady exponential wealth distribution, when preferential interactions are not considered. 

\subsection*{Nearest Neighbour Wealth Model}
In the initial preference, a single agent is chosen at random, while the second agent is randomly selected from those with wealth slightly below or above the first agent. 
Prior to each interaction, the wealth distribution is arranged in descending order, positioning the agent with the highest wealth at the top and the agents with the lowest wealth at the bottom.
Following this, a single agent is selected, and an interacting agent is randomly chosen from the adjacent agents positioned just below or above the selected agent in the sorted wealth distribution.
Furthermore, an extension to this model involves selecting the second agent randomly from a subset of N agents located below and above the initially chosen agent.

\subsection*{Delta Distance Wealth Model}
Here we modify the above preference by including the difference of wealth between two agents i,e. if the wealth difference between two agents is less than delta, then the interaction will happen.
Mathematically it can be describe as follows:
\begin{equation}\label{eq:pref1}
    |v-w| = \delta
\end{equation}
with $\delta$ being the constraint value. 
This simple selection criterion would be observed lead to a completely different wealth distribution depending on the values of $\delta$, as described in result section.
The motivation behind considering this kind of selection criterion is an observation that, in realistic society, often the interaction happens between the agents who belongs to similar state of wealth possession.


\section*{Boltzman equation} 

The Boltzman equation for a conservative wealth interaction, following the preference given by Eq.~(\ref{eq:pref1}), can be written as,
\begin{eqnarray}
    \frac{\partial f(v)}{\partial t} = \int_{w} \int_{w^{*}} \int_{v^{*}} \theta (|v^{*}-w^{*}|< \delta) f(v^{*})f(w^{*}) dv^{*} dw^{*} dw - \int_{w} \int_{w^{*}} \int_{v^{*}} \theta (|v-w|< \delta) f(v)f(w) dv^{*} dw^{*} dw
\end{eqnarray}
where the Heaviside Theta functions $\theta (|v^{*}-w^{*}|< \delta)$ and $\theta (|v-w|< \delta)$ takes care of the preferential wealth exchange. 
Also, as we are not considering debt in this work $\{v^{*}, w^{*}, w\}\in\mathcal{R}_{+}$.
Now the weak form of this equation can be written as:
\begin{align}
\begin{aligned}
    \frac{\partial}{\partial t}\int_0^\infty f(v)\phi(v)dv = \frac{1}{2}\int_{0}^\infty \int_{0}^\infty \theta(|v-w|< \delta)f(v)f(w)\\
    \left[\phi(v^{*})+\phi(w^{*})-\phi(v)-\phi(w)\right] dv dw
    \end{aligned}
\end{align}
Let, $\phi(v) = e^{-sv}$, So, $\phi(w) = e^{-sw}$ and similarly, $\phi(v^{*}) = e^{-sv^{*}}$ and $\phi(w^{*}) = e^{-sw^{*}}$. And from Eq.$(1)$ and Eq.$(2)$, we get, $\phi(v^{*}) = e^{-s\eta (v+w)}$ and $\phi(w^{*}) = e^{-s(1-\eta)(v+w)}$.
Putting this in Eq.$(5)$ and we get,
\begin{align}
 \begin{aligned}
\frac{\partial}{\partial t}\int_0^\infty f(v) e^{-sv} dv =\frac{1}{2}\int_{0}^{1} \int_{0}^\infty \int_{0}^\infty
\theta(|v-w|< \delta)f(v)f(w)\\
\left[e^{-s\eta (v+w)}+e^{-s(1-\eta)(v+w)}-e^{-sv}-e^{-sw}\right] dv dw d\eta
\end{aligned}   
\end{align}
Now by simplification we get as follows:
\begin{align}
    \begin{aligned}
\frac{\partial \Tilde{f}(s,t)}{\partial t} =\frac{1}{2}\int_{0}^{1} \int_{0}^\infty \int_{0}^\infty
\theta(|v-w|< \delta)f(v)f(w)
\left[e^{-s\eta (v+w)}+e^{-s(1-\eta)(v+w)}-e^{-sv}-e^{-sw}\right] dv dw d\eta \\
\implies \frac{\partial \Tilde{f}(s,t)}{\partial t}=\frac{1}{2}\int_{0}^{1} \int_{0}^\infty \int_{w-\delta}^{w+\delta}
\left[e^{-s\eta (v+w)}+e^{-s(1-\eta)(v+w)}-e^{-sv}-e^{-sw}\right]f(v)f(w) dv dw d\eta.
\end{aligned}
\end{align}
Now, when $\delta \longrightarrow \infty$, we get,
\begin{align}
\begin{aligned}
\frac{\partial \Tilde{f}(s,t)}{\partial t}=\frac{1}{2}\int_{0}^{1} \int_{0}^\infty \int_{0}^{\infty}
\left[e^{-s\eta (v+w)}+e^{-s(1-\eta)(v+w)}-e^{-sv}-e^{-sw}\right]f(v)f(w) dv dw d\eta.
\end{aligned}
\end{align}
This equation is known to give rise to a steady exponential distribution of wealth (\cite{Toscani}).
When we take the limit $\delta \longrightarrow 0$ we get,
\begin{align}
\begin{aligned}
\frac{\partial \Tilde{f}(s,t)}{\partial t}=\frac{1}{2}\int_{0}^{1} \int_{0}^\infty \left[e^{-s\eta (2w)}+e^{-s(1-\eta)(2w)}-2e^{-sw}\right]f^2(w) dw d\eta.
\end{aligned}
\end{align}
The above equation is difficult to tackle analytically, therefore we employ agent based numerical Monte-Carlo technique to compute the probability distribution function(PDF).
Results of the simulations are discussed in the following section.

\section*{Results: Numerical simulations}
\subsection*{Preference: leading to quasi-oligarchy}
\begin{figure}[H]
     \centering
      \includegraphics[scale=0.51]{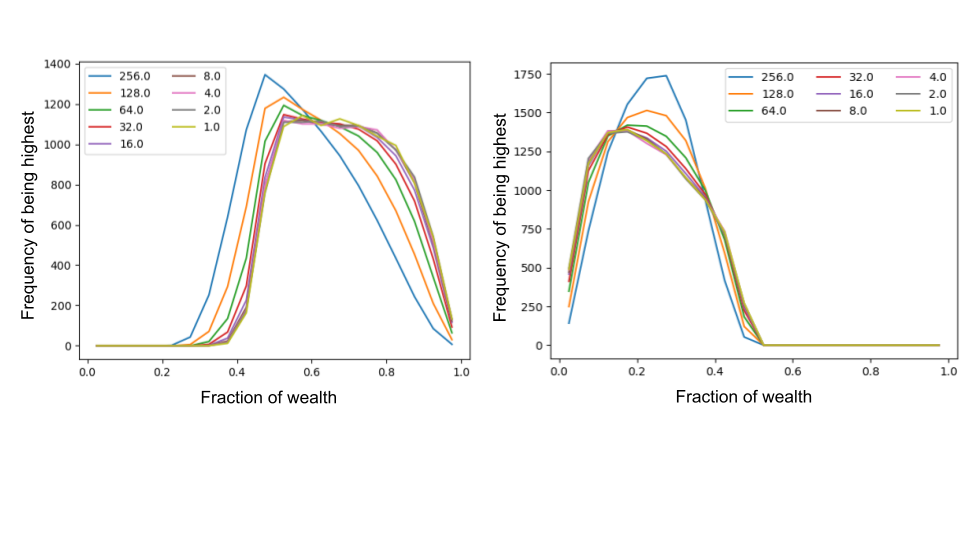}
   \caption{Distribution of wealth of the highest (left panel) and the second highest (right panel) wealth possessing agents for the nearest neighbor wealth model. 
   Different colors of the plot represent wealth distributions with varying sorting frequencies.}
     \label{fig:diff_sort}
  \end{figure}

Using the Nearest Neighbor Wealth Model (NNWM), we observe a concentration of wealth in the hands of a small number of agents. Our simulations encompassed a population of 256 agents, and we averaged the results over 1000 random realizations. While we did not observe a stable wealth distribution overall, we did identify a consistent distribution for the agent possessing the highest wealth.
We further investigated the impact of sorting frequency on the morphology of this steady wealth distribution. Fig.~\ref{fig:diff_sort} illustrates the wealth distribution of the highest wealth possessing agent for various sorting frequencies. Each color in the plot corresponds to a distinct scenario with different sorting frequencies. For instance, the blue curve represents the distribution when sorting occurs every 256 interactions, while the red curve corresponds to sorting every 32 interactions, and so on.
Additionally, the right panel of this figure presents the wealth distribution of the second-highest agent for different sorting frequencies. By examining these distributions, we gain insights into the effects of sorting frequency on the wealth distribution of both the highest and second-highest agents in the NNWM.

We see with increasing frequency of sorting the distribution spreads towards higher wealth in the left panel and towards lower wealth in the right panel.
This phenomenon suggests that as sorting occurs more frequently, the wealthiest agent tends to accumulate a greater share of wealth, while the second wealthiest agent experiences a corresponding decline in wealth. This pattern is indicative of a potential shift towards an oligarchic society.
However, our observations do not align with such a scenario. The reason behind this deviation can be attributed to the nature of our preference, wherein the interactions never cease, and wealth, no matter how minimal, is consistently redistributed. Consequently, the agent with the highest wealth at a given moment does not retain that position indefinitely. Instead, we witness a quasi-oligarchic behavior, characterized by a significant fraction of the total wealth being controlled by a select few individuals within the society. This differs from a true oligarchy, where all wealth is concentrated in the hands of a single individual.

Next, we present the wealth distribution obtained when the sorting frequency is set to one sorting per interaction, but the selection of the second agent is modified. Instead of choosing from the agents immediately above and below the first agent, we now randomly select the second agent from a set of 2N agents. This set is defined as a "Bound of neighbours" in our study.
As the bound of neighbour size (N) increases, we anticipate the wealth distribution to exhibit tendencies towards an exponential distribution. Conversely, as the bound size decreases, we expect to observe the quasi-oligarchic pattern discussed earlier.

\begin{figure}[H]
     \centering
      \includegraphics[scale=0.51]{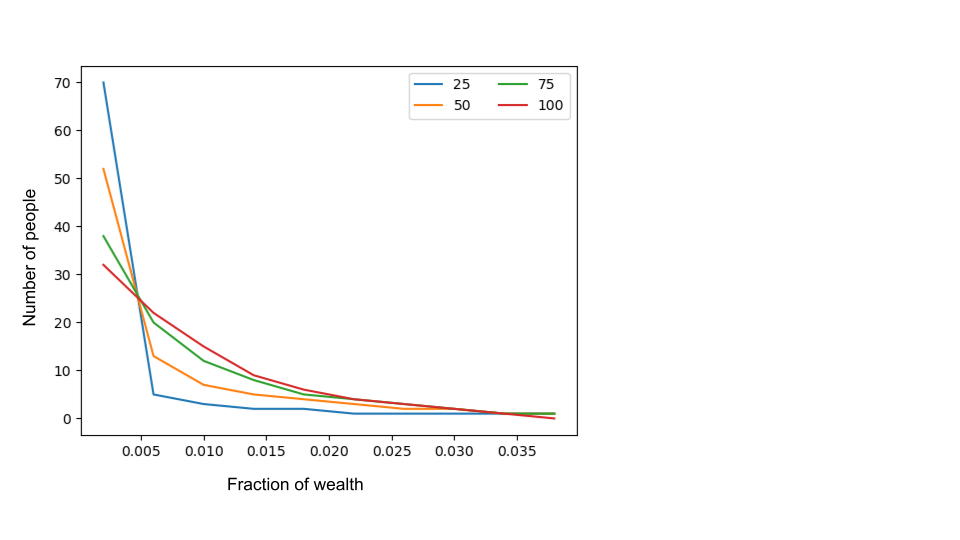}
   \caption{Wealth distributions for varying bound of neighbour size.}
     \label{fig:diff_cycle}
  \end{figure}

Fig.~\ref{fig:diff_cycle} illustrates the wealth distribution for different bound sizes, demonstrating the anticipated outcomes. It is important to note that no phase transitions are observed during the reduction of the bound size. Rather, we observe a continuous evolution of the wealth distribution's morphological pattern as the bound size varies.
Overall, our findings provide insights into the effects of bound size on wealth distribution within the kinetic exchange model.

\begin{figure}[H]
     \centering
      \includegraphics[scale=0.6]{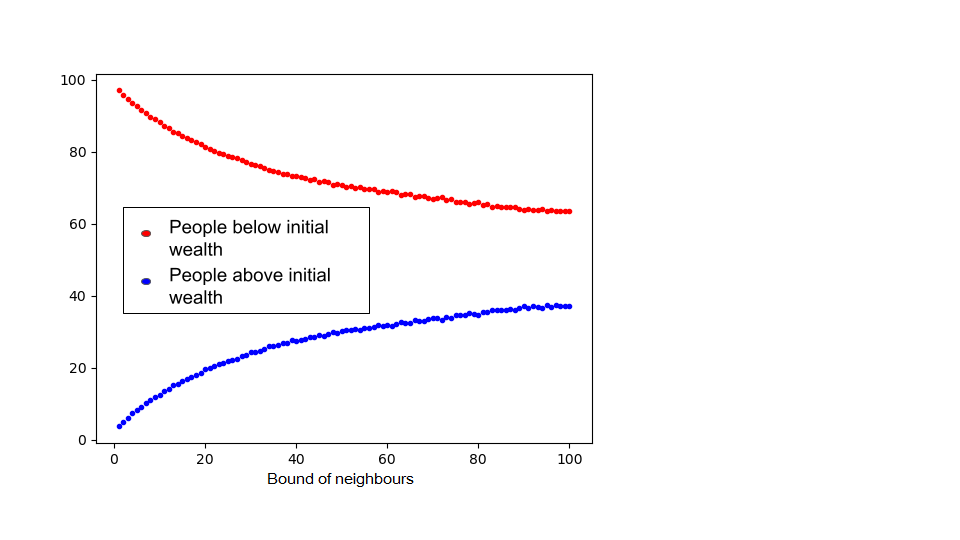}
   \caption{variation of the number of people having wealth below (red dots) and above (blue dots) their initial wealth after a certain number of interactions with the varying bound of neighbour. }
     \label{fig:diff_initial}
  \end{figure}

In fig.~\ref{fig:diff_initial}, we present findings that demonstrate the relationship between the number of individuals whose wealth surpasses or falls below their initial wealth with the number of bounds. Our observations reveal that as the bound size increases, there is a notable increase in the proportion of individuals who possess wealth greater than their initial wealth. This increment follows a power law pattern, suggesting a nonlinear relationship. Conversely, we also find that as the bound size increases, the number of individuals with wealth lower than their initial wealth decreases. These observations provide insights into the dynamics of wealth accumulation and distribution within the context of varying bound of neighbour size.
 
\subsection*{Preference: leading to segregation in a society}

\begin{figure}[H]
    \centering
    \includegraphics[scale=0.35]{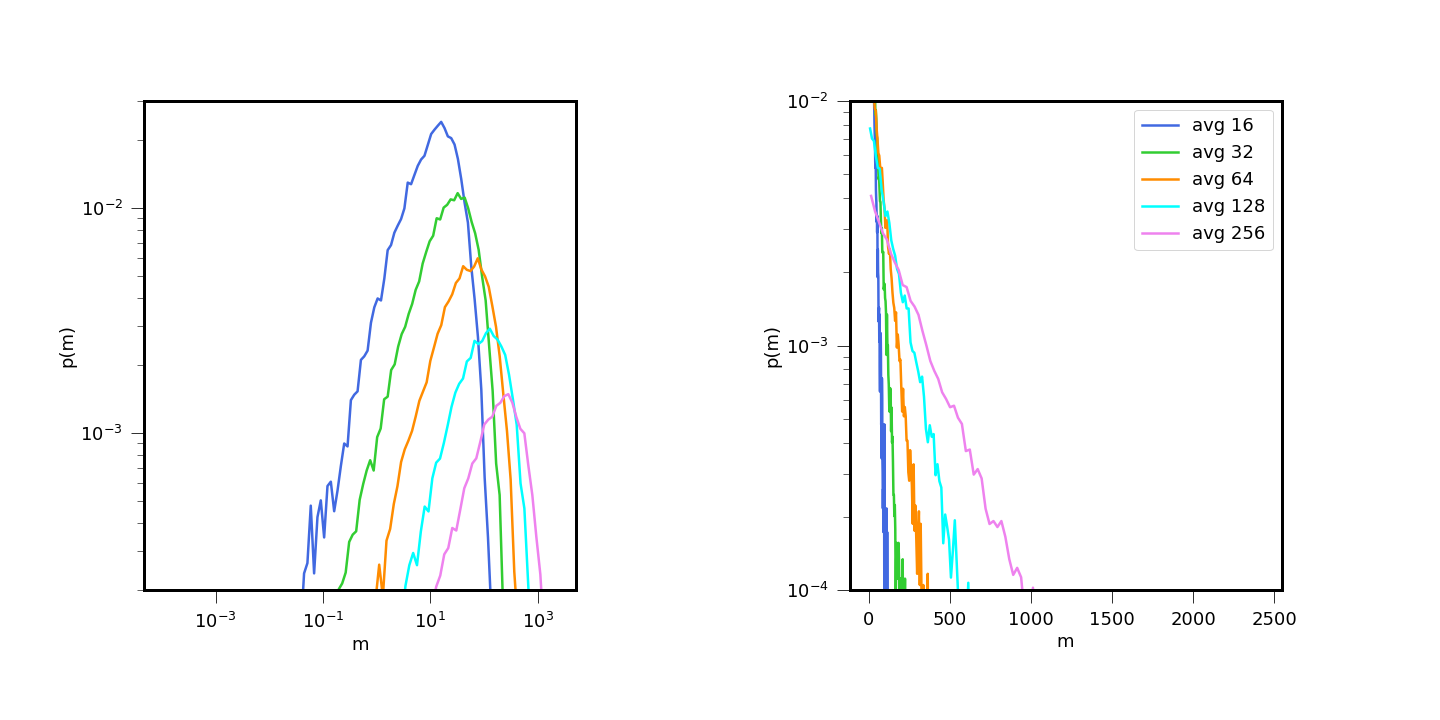}
    \caption{In the left panel, log-transformed of the simple exponential wealth distributions are shown for different average wealth on a log grid. We show the semi-log plot for the same distributions in the right panel. Different colors depict different average wealth of the distributions.}
    \label{fig:kin_model}
\end{figure}

Here in this section we proceed to showcase the results from the kinetic exchange simulation incorporating the preference.
To begin with, in Fig.~\ref{fig:kin_model} we show the log-transformed classical exponential wealth distribution governed by Eq.~(\ref{eq:dragulescu}) without the preferential interaction. 
The motivation to show this result is to demonstrate the capability of log-transformation, as we will see later, the log-transformation of the wealth distribution would be very helpful for the subsequent analysis. 
In the left panel of Fig.~\ref{fig:kin_model}, log-transformed wealth distribution for different average wealth is shown on a log grid, in the right panel we show the semilog plot for the same distributions.
While the linear trend of the distribution functions in the right panel validates the exponential nature of the wealth distribution, in the left panel the log-transformed distributions display peak structure. 
The position of the peak gives an idea about the average money in the system. 
For further validation we undertake the following exercise:
let us assume a random variable $(X)$ following the exponential distribution $(\lambda e^{-\lambda x})$. 
Further lets denote $Y = \ln{X}$ as the natural logarithm of the exponentially distributed random variable.
We wish to find the distribution for $Y$ and for this we begin with the cumulative distribution function (CDF) of $X$ as follows:
\begin{equation}\label{CDF}
P(Y \leq y) = P(\ln{X} \leq y) = P(X \leq e^{y})=\int_{0}^{e^{y}} \lambda e^{-\lambda x} dx = 1-e^{-\lambda e^{y}}.
\end{equation}
By differentiating Eq.~(\ref{CDF}), we get the probability distribution function of this transformed distribution as,
\begin{equation}\label{eq:log_t}
    f(y) = \lambda e^{(y-\lambda e^{y})}\,.
\end{equation} 
Taking natural logarithm of both the dependent and independent variable leads,
\begin{equation}\label{eq:loglog}
\ln{(f(\ln{y}))} = \ln{\lambda} + \ln{y} - \lambda e^{\ln{y}}
\end{equation}
This is the functional form for the distribution depicted in the left panel of Fig.~\ref{fig:kin_model}.
We find the maximum of this function corresponds to $y=\ln{\frac{1}{\lambda}}$, or the natural logarithm of the mean of the exponential distribution.

\begin{figure}[H]
    \centering
     \includegraphics[scale=0.85]{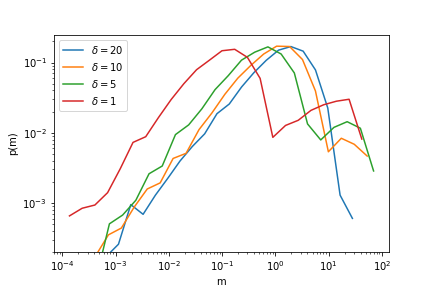}
   \caption{Wealth distribution with preference leading to segregation of the society. Different colors represent the variation of $\delta$ values.}
    \label{fig:diff_delta_dist}
 \end{figure}

In Fig.~\ref{fig:diff_delta_dist} we show the log transformed distribution function considering the preferential interaction condition with different $\delta$ (see Eq.~\ref{eq:pref1}). 
The simulation is performed with $10000$ number of agents and the distributions are shown after $10^{6}$ interactions.  
For larger $\delta$ values the distribution shows an expected exponential structure, but as the value of $\delta$ decreases the society eventually breaks into two groups with different mean wealth. 
We call the group with lower mean wealth as poor class and with higher mean wealth as the rich class.
Note that with decreasing $\delta$ the mean wealth of both the classes decrease, however this decrease is large for the poor class as compared to the other one. 

\begin{figure}[H]
     \centering
      \includegraphics[scale=0.85]{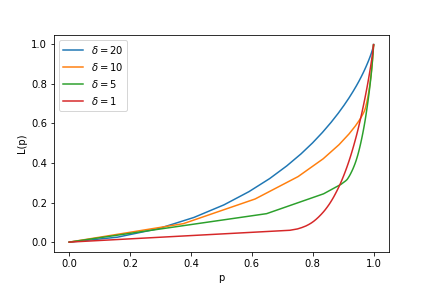}
   \caption{Lorenz curves for the wealth distributions shown in Fig.~\ref{fig:diff_delta_dist}.}
     \label{fig:diff_delta_Lorenz}
  \end{figure}

 In Fig~\ref{fig:diff_delta_Lorenz} we show the corresponding Lorenz curves for different $\delta$ values. 
 For higher $\delta$ the Lorenz curve shows a trend corresponding to the exponential wealth distribution. 
 As $\delta$ decreases the area under the Lorenz curve could be seen to decrease resulting in an increase in the overall wealth inequality. 
 With the decrease of $\delta$ the society evolves toward a state where a small fraction of agents ($\sim 10\%$ for $\delta=10$, $\gtrsim 10\%$ for $\delta=5$) posses a significant fraction of the entire wealth ($\gtrsim 40\%$ for $\delta=10$, $\gtrsim 70\%$ for $\delta=5$) in the system. 
 Further, with decreasing $\delta$ the fraction of agents possessing significant amount of wealth increases.

We also note that the wealth distribution with $\delta$ does not exhibit steady-state characteristics as seen in simple kinetic exchange interaction without preference. 
However, the distribution with different number of agents clearly imply that they are independent of that.
Here are these observations:

\begin{figure}[H]
     \centering
      \includegraphics[scale=0.51]{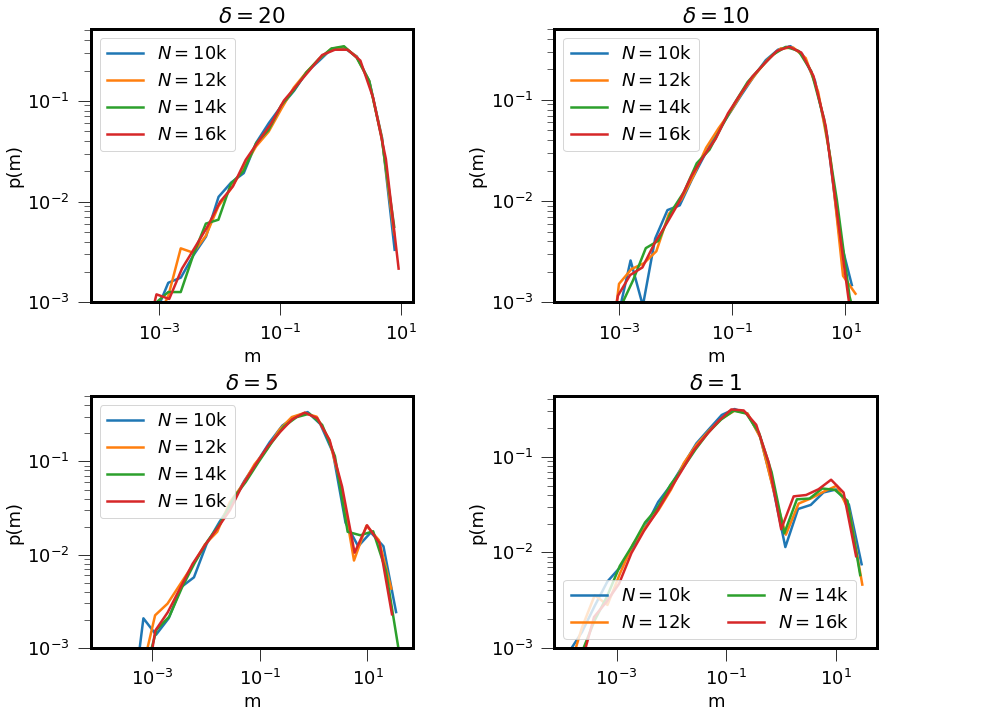}
   \caption{Distribution functions with different values of $\delta$ and different numbers of agent.
   The average wealth is chosen to be 1 to initiate the simulation. 
   The distribution functions can be observed to stay same with different number of agents.}
     \label{fig:diff_N}
  \end{figure}

In Fig.~\ref{fig:diff_N} we observe that the wealth distributions with this simple preference for different number of agents remain same inspite of being changing the values of $\delta$.

\begin{figure}[H]
     \centering
      \includegraphics[scale=0.51]{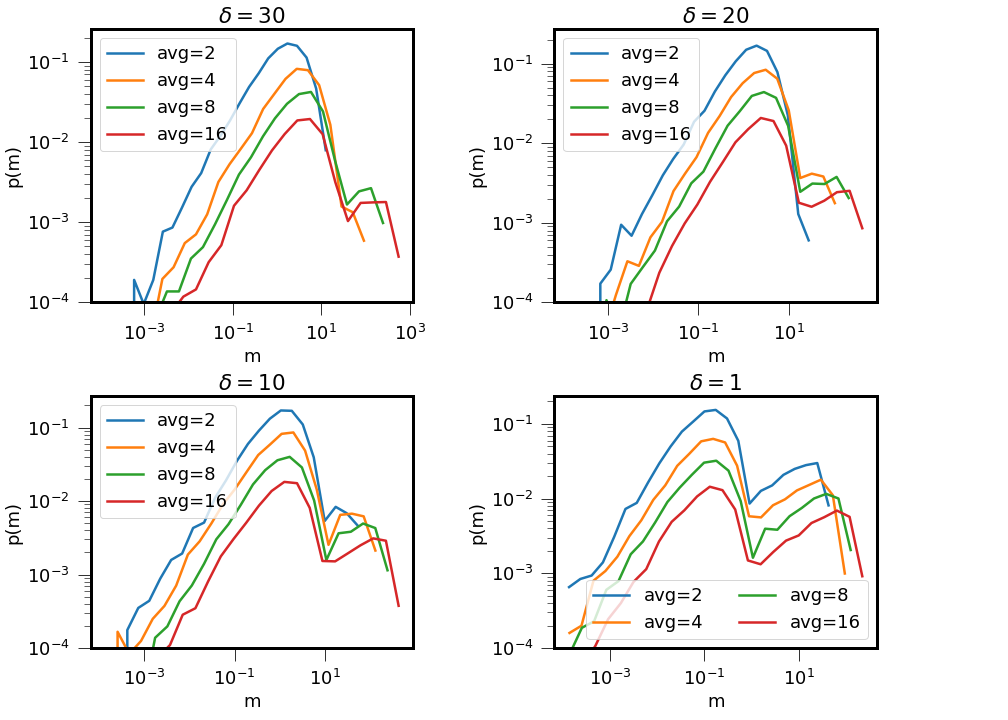}
   \caption{Distribution functions with different values of $\delta$ and different average wealth.
   }
     \label{fig:diff_avg}
  \end{figure}
  
Whereas Fig.~\ref{fig:diff_avg} shows that by changing the average money, we get different distributions although we consider the same value of $\delta$.
As a result, we can infer the following from these observations: while the wealth distribution is unaffected by the quantity of agents, it does alter with changes in the average income.

\section*{Conclusion}
Here in this work we present the role of preferential interaction in the context of kinetic exchange model of market. 
Such kind of model are realistically more favourable than the general kinetic exchange as usually economic agents prefer one kind of exchange than the other. 
We consider 2 preference scenarios in this work where agents are allowed to interact when they are neighbours in the wealth phase space. 
In the first preference, the first agent is selected randomly from the population. 
The second agent is then chosen randomly from a group of agents whose wealth levels are just below or above that of the first agent. 
This selection process allows for the consideration of agents with similar wealth levels in interactions.
Prior to each interaction, the wealth distribution is arranged in descending order, with the wealthiest agent positioned at the top and the least wealthy agents at the bottom. 
This arrangement ensures a clear hierarchy of wealth within the population.
With such a pilot scenario we observe the condensation of wealth in the hands of very few agents. 
Note that this is different from the scenario where the wealth condensates in the of only one agent, which has been explored in the earlier works (\cite{Chakrabarti,Boghusian}). 
As an extension to this model we adopt another scenario where in the post-sorting situation the second agent is choosen from the pool of 2N agents keeping the first agent in the middle. 
With such an extension we observe the number of agents having wealth more than their initial gets higher with higher value of N. 
The trend of such a process is observed to possess a power-law like behaviour. 

For the second preference we introduce selection criterion which stipulates that an interaction between two agents will only take place if the wealth disparity between them is less than or equal to a predefined threshold, $\delta$. 
By implementing this criterion, we simulate a series of interaction events among agents and observe the resultant wealth distributions.
We observe the emergence of segregation in the society due to such preferential interaction events. 
The segregated distribution does not attain a steady state and the segregation gets more prominent as we decrease the $\delta$ value. 

The two kind of preferences that we explored in this work bring the Kinetic Exchange Model to a situation which is more realistically closer as economic agents are usually observe to have a preference for every kind of realistic interaction scenario.
With the first preference, the Quasi-Oligarchic situation that we observed has been proven of economic relevance in earlier work \cite{Chakrabarti}, \cite{Boghusian}. 
The oligarchy that we observe is not a product of phase transition but a of a continuous dynamics.
The second preference led to society segregation which has been observed and reported in other works.
From this pilot study, we understand the origin of such segregation in economically relevant scenarios.
In future, we wish to explore the consequence of such oligarchy and segregation in the context of statistical physics more extensively.

\end{document}